\documentstyle[12pt,openbib,epsf]{article}

\begin{document}
\def\br{\begin{eqnarray}}
\def\er{\end{eqnarray}}
\def\be{\begin{equation}}
\def\ee{\end{equation}}
\def\a{\alpha}
\def\b{\beta}
\def\B{\Beta}
\def\d{\delta}
\def\D{\Delta}
\def\eps{\epsilon}
\def\g{\gamma}
\def\G{\Gamma}
\def\i{\int}
\def\l{\lambda}
\def\L{\Lambda}
\def\m{\mu}
\def\n{\nu}
\def\r{\rho}
\def\k{\kappa}
\def\e{\ell}
\def\({\left(}
\def\){\right)}
\def\s{\sigma}
\def\S{\Sigma}
\vspace{1cm}
\begin{center}
{\large{\bf Critical Coupling for Dynamical Chiral-Symmetry
Breaking with an Infrared Finite Gluon Propagator}}\\ 
A.~A.~Natale~\footnotemark
\footnotetext{e-mail: natale@axp.ift.unesp.br} 
and P.~S.~Rodrigues da Silva~\footnotemark\\
\footnotetext{e-mail: fedel@axp.ift.unesp.br}
Instituto de F\'isica Te\'orica, Universidade Estadual 
Paulista\\
Rua Pamplona, 145, 01405-900, S\~ao Paulo, SP Brazil
\end{center}
\thispagestyle{empty}
\vspace{1cm}

\begin{abstract}
We compute the critical coupling constant for the dynamical 
chiral-symmetry breaking in a model of quantum chromodynamics,
solving numerically the quark self-energy 
using infrared finite gluon propagators found as solutions
of the Schwinger-Dyson equation for the gluon, and
one gluon propagator determined in numerical lattice 
simulations. The gluon mass scale screens the force
responsible for the chiral breaking,
and the transition occurs only for a
larger critical coupling constant than the one obtained with the
perturbative propagator. The critical coupling shows a great
sensibility to the gluon mass scale variation, as well as
to the functional form of the gluon propagator.
\end{abstract}

\newpage
\section{Introduction}

\noindent
The idea that quarks obtain effective masses as a result of a 
dynamical breakdown of chiral symmetry (DBCS) has received a
great deal of attention in the last years~\cite{nambu,farhi}.
One of the most common methods used to study the quark mass
generation is to look for solutions of the Schwinger-Dyson
equation for the fermionic propagator. It is known that 
above a certain critical coupling $(\a_c \equiv g_s^2/4\pi)$
a nontrivial self-energy solution bifurcates away from the
trivial one. Numerical evaluation of this critical coupling
in QCD with three and four flavors gives 
$\a_c \sim {\cal O} (1)$~\cite{atkinson,roberts}.

Parallel to the study of DBCS a lot of effort has also been
done to obtain the nonperturbative behavior of the gluon
propagator~\cite{corn,stingl,cudell}, and perhaps one of the
most interesting results is the one where it is argued that 
the gluon may have a dynamically generated mass~\cite{corn}.
The study of the infrared behavior of the gluon propagator
was also performed numerically on the lattice~\cite{bernard},
and more recent numerical simulation give strong evidence
for an infrared finite gluon propagator in the Landau 
gauge~\cite{bernard2}. It is worth mentioning that from the
phenomenological point of view, the existence of a ``massive
gluon" may shed light on several reactions where long 
distance QCD effects can interfere, and examples of the 
possible consequences can be found in the literature, see,
for instance, Ref.~\cite{parisi,halzen,field}. 

Much work has yet to be done about the infrared behavior 
of the gluon propagator, but it is clear that its 
implications have to be tested in all possible problems.
It is possible that the constraint coming from DBCS, and
other phenomenological studies~\cite{parisi,halzen,field}
will provide a map of the infrared gluon propagator.
Since the bifurcation point for DBCS was studied up to
now with the perturbative $1/k^2$ gluon propagator, it
is natural to ask what is going to happen with the 
infrared finite propagators that have been found through
solutions of the gluonic Schwinger-Dyson equation or
using Monte Carlo methods, and, moreover, to look for
the consequences of different forms of non-perturbative
gluon propagators. It is intuitive that the
force necessary for condensation is going to be 
screened if the gluon propagator is infrared finite,
therefore, the actual critical coupling constant should
be larger, and this is what we will investigate in this
work.

We will present the Schwinger-Dyson equation of our 
problem, and first we will discuss the critical coupling
for the linear approximated problem in the case of a bare
gluon mass. This will teach us on the general behavior of
the critical coupling constant as a function of the gluon
mass. Secondly, we perform a numerical calculation of the
full nonlinear equation, for two different gluon propagators
resulting from solutions of the gluon polarization tensor, 
and one obtained from numerical simulation on the lattice.
In the conclusions we discuss the differences in the
critical coupling for each of the ``massive'' gluon
propagators, arguing that its value definitively gives
information about the infrared gluon propagator.

\section{Quark propagator Schwinger-Dyson equation}

The Schwinger-Dyson equation for the quark propagator 
in Minkowski space is
\be
S^{-1}(p)=\not\!{p}
-\imath \frac{4}{3} \i \frac{d^4q}{(2 \pi )^4}
\g_\m S(q)\G_\n(p,q)g^2D^{\m\n}(p-q) ,
\label{sd}
\ee
where we write the gluon propagator in the form
\be
g^2D^{\m \n}(q)= \frac{4\pi \a(-q^2/\L^2)}{q^2}
\( -g^{\m \n}+\frac{q^{\m}q^{\n}}{q^2} \).
\label{gau}
\ee
The propagator has been written  
in the Landau gauge, which will be used throughout our work.
In the above equations $\G_\n(p,q)$ is the vertex function, and
$\a(-q^2/\L^2)$ is the QCD running coupling constant, for
which we know only the ultraviolet behavior, and to solve
Eq.(1) we make the same ansatz of 
Ref.~\cite{atkinson,roberts} 
about its behavior for the full momentum scale
\be
\a(-q^2/\L^2)= \frac{12\pi/(33-2n_f)}{\ln(\tau + 
\frac{-q^2}{\L^2})}.
\label{alfa}
\ee
Eq.(3) goes continuously to the perturbative result, and has
already been used in phenomenological applications.

To proceed further we also need to introduce an ansatz for
the quark-gluon vertex $\G^\m(p,q)$, which must satisfy a
Slavnov-Taylor identity that, when we neglect ghosts, reads
\be
(p-q)_\m \G^\m(p,q)=S^{-1}(p)-S^{-1}(q).
\label{st}
\ee
This identity constrains the longitudinal part of the 
vertex, and if we write $S^{-1}(p)$ in terms of scalar
functions
\be
S^{-1}(p)=A(p) \not\!{p} - B(p), 
\label{fpro}
\ee
we find the solution~\cite{krein}
\br
\G^\m(p,q) &=& A(p^2)\g^\m+\frac{(p-q)^\m}{(p-q)^2}
\([A(p^2)-A(q^2)] \not\!{q} - 
[B(p^2)-B(q^2)] \) \nonumber \\ 
&& +  \,\, transverse \, part,
\label{vert}
\er
which is a much better approximation than the use of the bare 
vertex. Assuming that the transverse vertex part vanishes 
in the Landau gauge we obtain
\be
D^{\m\n}(p-q)\G_\n(q,p)=D^{\m\n}(p-q)A(q^2)\g_\n,
\label{fin}
\ee
and arrive at the approximate Schwinger-Dyson equation
\be
[A(p^2)-1]\not\!{p}-B(p^2) = \imath\frac{4}{3} \i \frac{d^4q}{(2 \pi 
)^4} g^2 D^{\m\n}(p-q)\g_\m \frac{A(q^2)}{A(q^2)\not\!{q}-B(q^2)}
\g_\n.
\label{v2}
\ee
Going to Euclidean space, we will be working with the following
nonlinear coupled integral equations for the quark wave-function
renormalization and self-energy
\br
[A(P^2)-1]P^2 &=& \frac{16\pi}{3} \i \frac{d^4Q}{(2\pi)^4}
\frac{\a((P-Q)^2/\L^2)}{\Phi[(P-Q)^2]} \nonumber \\
&& \times \( P.Q +2 \frac{P.(P-Q)Q.(P-Q)}
{(P-Q)^2}\) \nonumber \\
&& \times \frac{A^2(Q^2)}{A^2(Q^2) Q^2+B^2(Q^2)},
\label{Aeq}
\er
\be
B(P^2)=16\pi \i \frac{d^4Q}{(2\pi)^4}
\frac{\a((P-Q)^2/\L^2)}{\Phi[(P-Q)^2]}
\frac{A(Q^2)B(Q^2)}{A^2(Q^2)Q^2+B^2(Q^2)},
\label{Beq}
\ee
where $Q^2=-q^2$ and $P^2=-p^2$, and we introduced a function
$\Phi[(P-Q)^2]$ which, in the case of the perturbative
propagator, is simply $\Phi[(P-Q)^2]=(P-Q)^2$, for the
massive bare gluon it will have the form 
$\Phi[(P-Q)^2]=(P-Q)^2+m_g^2$, and will be a more
complex expression in the case of a dynamically generated
mass.

\section{The linear problem with a massive bare gluon}

Eq.(\ref{Aeq}) and Eq.(\ref{Beq}) possess the trivial
solution $A(P^2)=1$ and $B(P^2)=0$ for small values of
the coupling constant. We can also see that $B(P^2)$ 
depends on $B(P^2)$ at first order, whereas $A(P^2)$
has a higher order dependence on $B(P^2)$. In order
to examine the possibility that a nontrivial solution,
$B(P^2)$, branches away from the trivial one at a
critical coupling, $\a_c$, we examine the so-called
bifurcation equation~\cite{pimbley}. This involves
differentiating Eq.(\ref{Aeq}) and Eq.(\ref{Beq})
functionally with respect to $B$ and then setting
$B=0$. Since the equation for $A(P^2)$ depends at
least quadratically on $B(P^2)$ it will be droped
at leading order from the bifurcation problem, and 
we substitute $A(P^2)$ by $1$ in the bifurcation
equation that will come out from Eq.(\ref{Beq}).
We will deviate from the standard bifurcation theory
proceeding as in Ref.~\cite{gusynin}, and instead
of substituting $Q^2+B^2(Q^2)$ by $Q^2$ in the
denominator of Eq.(\ref{Beq}), we will replace this
term by $Q^2+\d B^2(0)$ and define the dynamical 
fermion mass $(m_f)$ by the normalization condition
\be
\d B(0)=m_f.
\label{dma}
\ee
We finally arrive at our bifurcation equation
\br
\d B(P^2)&=& \frac{16\pi}{(2\pi)^3} \i dQ^2 \, \i d\theta
\sin^2\theta \frac{Q^2}{Q^2+m_f^2} \nonumber \\
&& \times \frac{\a[(P-Q)^2/\L^2]}{(P-Q)^2+m_g^2} \d B(Q^2),
\label{bifeq}
\er
where we have already assumed a bare massive gluon.

Our main intention in this section is to verify the 
gross behavior of the critical coupling constant
with the existence of an infrared finite gluon
propagator, this is the reason for having selected
a bare massive gluon in Eq.(\ref{bifeq}). The
details of a dynamically generated gluon mass will
be left for the next section. Eq.(\ref{bifeq})
is a standard Fredholm equation with a positive
kernel, and, requiring $\d B(P^2)$ to belong to
$L^2$, the spectrum is discrete with a smallest
value $\a_c$ such that we have the trivial solution
$\d B(P^2)\equiv 0$ for $0 < \a < \a_c $, and the
nontrivial one if $\a \geq \a_c$.

We can still make some simplifier approximations
before estimating $\a_c$, making the following
substitutions
\be
\a((P-Q)^2/\L^2) \rightarrow \theta(P^2-Q^2)\a(P^2/\L^2)
+\theta(Q^2-P^2)\a(Q^2/\L^2),
\label{anga}
\ee
and
\be
\frac{1}{(P-Q)^2+m_g^2}=\frac{1}{P^2+m_g^2}\theta(P^2-Q^2)
+\frac{1}{Q^2+m_g^2}\theta(Q^2-P^2),
\label{angp}
\ee
which is known as the angle approximation, and introduces
an error of about $10 \% $ in the calculation~\cite{roberts}.
Defining the variables $ x= P^2 /m_f^2 $, $ y= Q^2 /m_f^2 $,
$\e=\L^2/m_f^2$, $\k=m_g^2/m_f^2$, and 
$f(P^2)=\d B(P^2)/m_f$, we obtain 
\be
f(x)=\frac{1}{\pi} \i dy \, K(x,y) \, f(y),
\label{fredeq}
\ee
where
\be
K(x,y)= \frac{\a(x/\e)}{x+\k} \frac{y}{y+1} \theta(x-y)
+\frac{\a(y/\e)}{y+\k}\frac{y}{y+1}\theta(y-x).
\label{kernel}
\ee

The kernel $K$ is square integrable
\br
\parallel K \parallel^2 &=& \i_0^\infty dx \, \i_0^x dy \,
\frac{\a^2(x/\e) y^2}
{(x+\k)^2(y+1)^2} \nonumber \\ 
&& + \i_0^\infty dx \, \i_x^\infty dy \,
\frac{\a^2(y/\e) y^2}{(y+\k)^2(y+1)^2},
\label{modk}
\er
therefore Eq.(\ref{fredeq}) has a nontrivial $L^2$ solution
for $\a_c$ on a point set. The smallest eigenvalue ($\a_c$)
for which Eq.(\ref{fredeq}) has a nontrivial square integrable
solution, is the first bifurcation of the nonlinear equation,
and satisfy
\be
\frac{1}{\pi} \, \parallel K \parallel \,
= \, 1.
\label{cond}
\ee
Table 1 gives the critical value $\a_c$ as a function
of $\e$ and $\k$.
\begin{table}
\begin{center}
\begin{tabular}{|l|c|r|}\hline\hline
%\multicolumn{3}{|c|}\hline\hline
\textbf{$\ell$} & \textbf{$\k$} & \textbf{$\alpha_c$} \\ \hline\hline
$10^4$ & 1 & 0.6971 \\ \hline
$10^4$ & $10^2$ & 0.9440 \\ \hline
$10^4$ & $10^3$ & 1.4853 \\ \hline
$10^6$ & $  1 $ & 0.5568 \\ \hline
$10^6$ & $10^2$ & 0.6607 \\ \hline
$10^6$ & $10^4$ & 0.9489 \\ \hline
$10^{10}$ & $10^4$ & 0.6226 \\ \hline
$10^{10}$ & $10^6$ & 0.7822 \\ \hline
$10^{10}$ & $10^8$ & 1.2278 \\ \hline
\end{tabular}
\end{center}
\caption{Critical coupling constant $(\a_c)$ as a function of 
$\e=\L^2/m_f^2$, and $\k=m_g^2/m_f^2$ for $n_f=4$.}
\label{table1}
\end{table}

The values of Table 1 were obtained with
$n_f=4$ but they do not change appreciably
as we change $n_f$.
As could already be seen in Eq.(\ref{modk}), 
if we increase the gluon masses 
we can satisfy Eq.(\ref{cond}) only with
larger critical coupling constants, and this is what
we can expect from the numerical solution of the 
complete nonlinear problem. We stress that the
values in Table 1, which are approximate solutions
of Eq.(\ref{cond}),
can only give a qualitative idea of the problem,
because the actual solution, without the many
simplifications performed until we arrived at 
Eq.(\ref{modk}), will obviously 
differ from the results of Table 1.

\section{The critical coupling for infrared finite
propagators}

In this section we solve Eq.(\ref{Aeq}) and
Eq.(\ref{Beq}) numerically without further
approximations. The numerical code we used is the
same of Ref.~\cite{williams}, and the criterion to
determine the critical coupling is the one of
Ref.~\cite{roberts}. With the perturbative
gluon propagator we obtain (with $n_f=4$)
\be
\a_c=  0.854,
\label{apert}
\ee
which is compatible with the calculations of 
Ref.~\cite{atkinson,roberts}. We
will solve the gap equations with three different
propagators which we discuss in the sequence.

One of the infrared finite propagators found in
the literature was determined by Cornwall~\cite{corn}
\be
\Phi (Q^2)=D_c^{-1}(Q^2)= [Q^2 + m_g^2(Q^2)] bg^2 
\ln[\frac{Q^2 + 4m_g^2(Q^2)}{\L^2}], 
\label{propc}
\ee
where $m_g^2(Q^2)$ is the momentum-dependent dynamical 
gluon mass
\be
m_g^2(Q^2)= m_g^2 \left[ 
\frac{\ln\left( \frac{Q^2 + 4m_g^2}{\L^2} \right)}
{\ln{\frac{4m_g^2}{\L^2}}} \right]^{-12/11},
\label{mcor}
\ee
$g^2\sim1.5-2$ is the strong coupling constant, and
$b=(33-2n_f)/48\pi^2$ is the leading order coefficient
of the $\b$ function of the renormalization group equation.
This form for the propagator was obtained as a fit to
the numerical solution of a gauge invariant set of 
diagrams for the gluonic Schwinger-Dyson equation.

Another infrared finite gluon propagator has been
found by Stingl and collaborators~\cite{stingl}. Its
form agrees with that derived by Zwanziger based
on considerations related to the Gribov 
horizon~\cite{gribov}, and is given by
\be
\Phi(Q^2)=D_s^{-1}(Q^2)= Q^2 + \m^4/Q^2 ,
\label{propst}
\ee
where $\m$ is a scale not determined in 
Ref.~\cite{stingl}. It is interesting to note that
the Bernard {\it et al.}~\cite{bernard2} lattice
result for the gluon propagator can be fitted by
Eq.(\ref{propc}) as well as Eq.(\ref{propst}).
These propagators, apart from some multiplying constant,
approach the perturbative gluon propagator in the
small mass limit.

Finally, Marenzoni {\it et al.}~\cite{bernard2}
also performed a lattice study of the gluon 
propagator in the Landau gauge, obtaining for its
infrared behavior the following fit
\be
\Phi(Q^2)=D_m^{-1}(Q^2)= m_g^2 + Z  Q^2 (Q^2/\L^2)^{\eta},
\label{propl}
\ee
where $m_g, Z$ and $\eta$ are constants determined
with the numerical simulation. $m_g$ is of ${\cal O}(\L)$, 
$Z \simeq 0.4$ and $\eta \simeq 0.5$, what is 
slightly different from the previous propagators.
The results of Bernard {\it et al.} also show the 
behavior $(Q^2)^\eta$, but with a smaller value for
$\eta$.

With the above propagators we computed the dynamical 
fermion mass as a function of the coupling constant.
As in Ref.~\cite{roberts} the results were fitted
by a function
\be
h(\a) = \b (\a-\a_c)^\g,
\label{acrit}
\ee
characteristic of a phase transition phenomena.
We have not found large differences in the values of
the critical coupling as we variated the number
of fermions, therefore, the fitting will be presented
for $n_f=4$. In Fig.1 we plot $-1/\ln{B(0)}$ as a function
of the coupling constant, for the Cornwall propagator (see 
Eq.(~\ref{propc})). The curves in Fig.1 were obtained
for $m_g = 2 \L$ and $m_g = 2.2 \L$, and as expected
from the example of the previous section if we increase
the gluon mass the critical coupling also increases.
These gluon masses are consistent with the values
determined phenomenologically in the last work of
Ref.~\cite{corn}.
The parameters of Eq.(\ref{acrit}) and the critical
coupling are given by
\br
\b=1.0785, \;\; \g=0.2535, \;\; 
\a_c=0.8692, \;\; (m_g=2.0\L); \\
\b=0.8424, \;\; \g=0.2682, \;\; 
\a_c=1.4211, \;\; (m_g=2.2\L).
\label{curvacor}
\er
As will become clear in the following, not only the
value of the gluon mass scale is important to characterize
the phase transition, but the precise 
form of the gluon propagator
will affect considerably the value of the critical coupling.
In this case, as well as in the next ones, we verified that
for small gluon masses we start having dynamically generated
quark masses for critical couplings quite close to the
value obtained with the $1/Q^2$ propagator (see Eq.(25)).
After some value of the gluon mass the critical coupling
deviates very fast from the value of Eq.(19). An example of
this behavior is shown in Eq.(26), where the coupling constant
is almost twice the value of Eq.(25), although we obtained it
increasing the previous gluon mass value only by $ 10 \% $!
Finally, the
chiral symmetry breaking for this propagator was also
studied in Ref.~\cite{haeri} with different results.
The main difference lies in the fact that in Ref.~\cite{haeri}
it is assumed a complete cancellation between the coupling
constant of the vertex function and the coupling in the
denominator of Eq.(\ref{propc}). Therefore, the mass
generation in Ref.~\cite{haeri} does not depend at
all on the coupling constant, and it is far from 
clear to us that such cancellation should be performed.

Using the propagator determined by Stingl and 
collaborators~\cite{stingl}, we obtain the curves
shown in Fig.2 for $\m^2=0.25\L^2$ and $\m^2=0.30\L^2$,
and described by Eq.(\ref{acrit}) with
\br
\b=0.2482, \;\; \g=0.4784, \;\; 
\a_c=2.9038, \;\; (\m^2=0.25\L^2); \\
\b=0.2946, \;\; \g=0.3362, \;\; 
\a_c=6.4720, \;\; (\m^2=0.30\L^2).
\label{curvast}
\er
Note that the values for the critical coupling constants
are quite large. We rely on these numbers based on the
continuous growth of the coupling constant from a value
near the one of Eq.(19) for small gluon masses, to the
ones of Eq.(27) and (28) as the mass is increased. It 
is known for several other theories with chiral breaking
for coupling constants of $\cal{O}$$(1)$, that higher order
corrections do not modify the critical behavior shown by
the ladder approximation~\cite{uma}, and we expect the 
same to hold here. Comparing Fig.2 to Fig.1
we see that the dynamically generated mass is much
smaller for this propagator, than with the
Cornwall one. Performing the calculation
for $\m\approx$$\, \cal{O}$$(3.0 \L)$ we do not obtain a
significative signal of chiral symmetry breaking, i.e.
if there is a dynamical mass it is much smaller
than $\L$, and do not satisfy our numerical criterion to
recognize mass generation~\cite{roberts}. This result
is compatible with the one of Ref.~\cite{hawes}, where it
was verified that the quark condensate is consistent
with zero above a certain critical value of $\m$ for
this same gluon propagator. Here we foresee a problem
for the Stingl {\it et al.}~\cite{stingl} propagator,
because as shown by Cudell and Nguyen~\cite{halzen}
we need $\m\approx$$\, \cal{O}$$(3.0 \L)$ to obtain a
correct description of diffractive scattering with
this propagator. We stress that not only the gluon
mass scale is important to determine the critical
coupling constant, but also the functional form
of the propagator.

Fig.3 contains the critical curve for the lattice
propagator (Eq.(\ref{propl})) in the case of
$m_g=0.7\L$, and with
\be
\b=0.4588, \;\; \g=0.2870, \;\; \a_c=3.9712, \;\; (m_g=0.7\L).
\label{clat}
\ee
The Marenzoni {\it et al.} propagator gives 
a value for the critical coupling constant which
is intermediate between the other two propagators
that we discussed up to now. If we increase the gluon
mass we will also find a point where the symmetry
is not broken anymore, however, this will occur for
larger masses than the one predicted in 
Ref.~\cite{bernard2} $(m_g \approx \L)$.
Comparing all the results
it becomes clear that, for masses of the same
order, the softer is the
propagator in the infrared the larger will
be the critical coupling for chiral symmetry
breaking. 

It is known from early studies of DBCS with the
perturbative $(1/k^2)$ gluon propagator~\cite{peskin},
that the chiral symmetry is broken when the product of the
coupling constant $(\a)$ with a Casimir eigenvalue $(C_F)$,
which depends on the fermion representation, is larger than
a certain critical value. With the introduction of a gluon
mass scale this is obviously not the case. We have found
that the gluon mass scale plays an important role in this
mechanism, the larger it is the stronger must be the coupling
constant to generate dynamical quark masses. In this way, 
there is some chance that the single-gluon-exchange approximation,
as presented here, makes no sense to obtain consistent solutions 
of Schwinger-Dyson equations, except for very small gluon masses,
or for theories with fermionic representations condensating in
a channel with large eigenvalues $C_F$. For large gluon masses
and fundamental representation quarks the critical coupling
becomes quite large, and it is more likely that we should
consider some effect due to confinement, which could be
responsible for a correct treatment of DBCS.

In a related work Papavassiliou and Cornwall~\cite{papa}
also considered the gap equation with massive gluons
coupled to the vertex equation. They found that non-singular
solutions of the vertex equation requires large gluon masses,
which erase the chiral symmetry breaking! In our calculation
we introduced a phenomenological coupling constant that
freezes at low momentum depending on the parameter $\tau$
of Eq.(\ref{alfa}). This parameter is determined when
we find the critical coupling $\a_c$ for each gluon mass.
Both works can be related if we choose $\tau \propto
m^2_g / \L^2$. With this choice when we increase the gluon
mass the coupling decreases, and we may never have a
coupling constant large enough to generate chiral symmetry
breaking, arriving at the same inconsistency found in
Ref.~\cite{papa}. We stress that the determination of the 
critical coupling for chiral symmetry breaking, and the
verification of the freezing of the coupling constant as
predicted in Ref.~\cite{papa}, are crucial tests for the
existence of a gluon mass scale, as well as can be a good
indicator of the true infrared gluon propagator.   

\section{Conclusions}

We studied the critical coupling constant for the dynamical
chiral-symmetry breaking in a model of quantum chromodynamics,
using infrared finite gluon propagators found as solutions of
the Schwinger-Dyson equation for the gluon, as well as one
gluon propagator determined in numerical lattice simulations.
We first calculated the eigenvalue condition for the
linear bifurcation equation of the quark self-energy, finding
that a bare gluon mass scale screens the force responsible for
the chiral breaking, and the transition occurs at a larger
critical coupling constant if we increase the ratio of the
gluon to fermion mass. Secondly, we solved numerically
the full quark self-energy equation for some infrared 
finite gluon propagators. The result confirm our linear
approximation, indicating that as we increase the gluon
mass the critical coupling constant will be larger.
We also verified that the functional form of the propagator
is also important to characterize the chiral transition.

With the Cornwall propagator (Eq.(\ref{propc})) and gluon
masses of the order that are expected phenomenologically,
we obtain critical coupling constants not far away from
the one obtained with the $1/k^2$ propagator. For this
propagator our calculation differs from the one of
Ref.~\cite{haeri} by the reason explained in Section 4.
With the Stingl {\it et al.} propagator (Eq.(\ref{propst}))
the chiral transition will occur only for quite large
values of the coupling constant. If the gluon mass 
scale is of {\cal O}$(\L)$ the critical coupling is
one order of magnitude larger than the one obtained
with the perturbative propagator. Unfortunately, a
phenomenological study of diffractive scattering
with the Stingl propagator demand gluon masses of
${\cal O}(3\L)$, for which there is no symmetry
breaking! This means that this propagator does not
represent the actual gluon infrared behavior,
or the model of diffractive scattering of Ref.~\cite{halzen}
must be modified. The 
Marenzoni {\it et al.} propagator leads
to a picture of the chiral transition that is consistent
phenomenologically, but with a larger value for the
critical coupling constant. In general, the softer is
the propagator in the infrared the larger will be the
critical coupling. As discussed at the end of last
section, for large gluon masses, it is possible
that confinement effects cannot be discarded, due to
the large critical coupling constants involved
in these cases.
The value of the critical coupling constant can
be used as a tool to study the infrared behavior of
the gluon propagator, and associated with other
phenomenological calculations (like the ones of
Ref.~\cite{parisi,halzen,field}) may provide a
map of the gluon propagator for every momenta
scale. 

\section*{Acknowledgments}

We have benefited from discussions with G.~Krein and J.~Montero.
This research was supported in part by the Conselho Nacional
de Desenvolvimento Cientifico e Tecnologico (CNPq) (AAN), 
and in part by Fundacao de Amparo a Pesquisa do
Estado de Sao Paulo (FAPESP) (PSRS).

\newpage
\begin{figure}[htb]
\epsfxsize=0.8\textwidth
\begin{center}
\leavevmode
\epsfbox{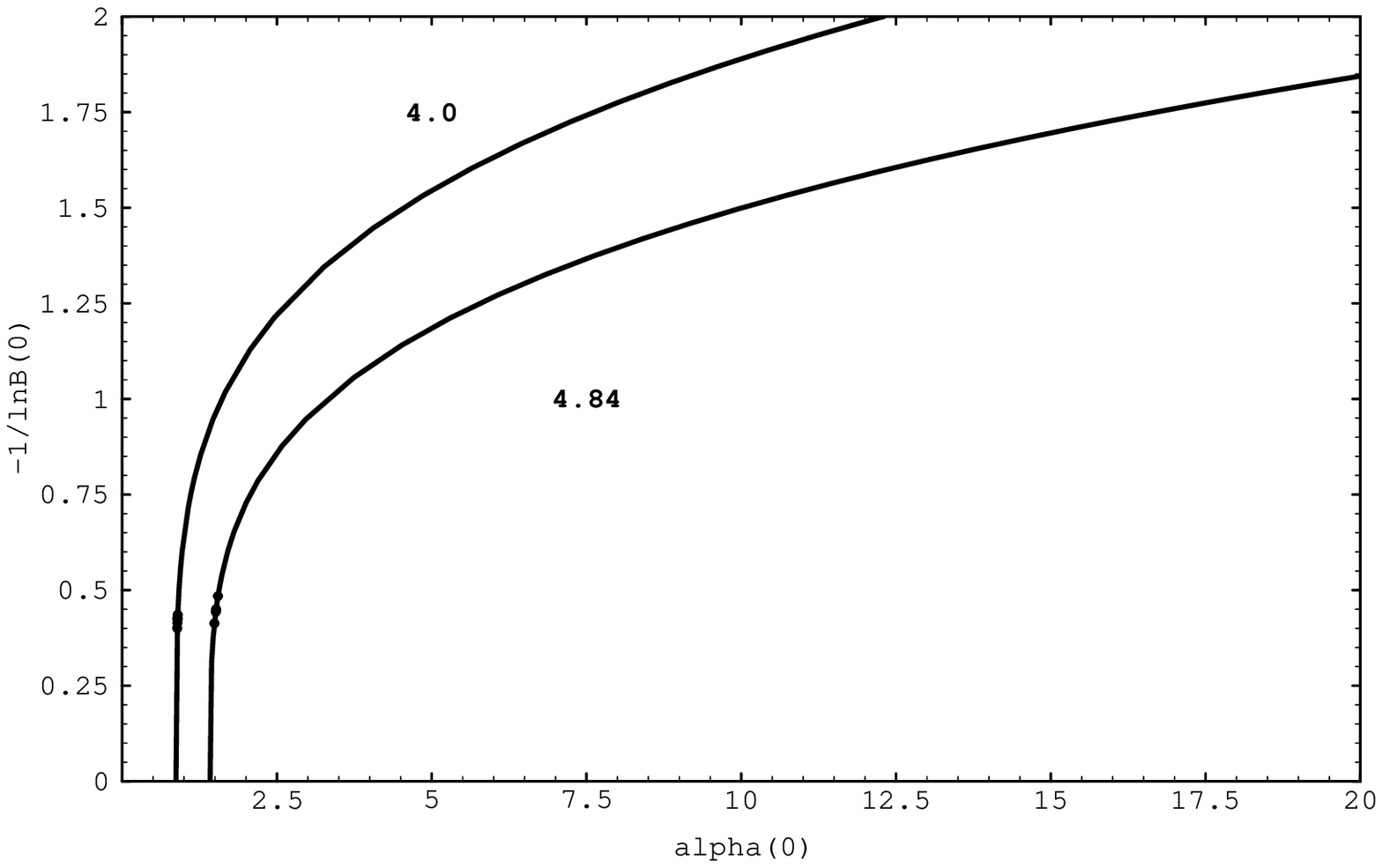}
\end{center}
\caption{Results of the evaluation of Eq.(\ref{Beq}) as
a function of the coupling constant with the propagator
of Eq.(\ref{propc}). We show some of the calculated points, and
the curve is the result of the fitting by Eq.(\ref{acrit}). The
calculation was performed for $n_f=4$, the upper curve is
for $m^2_g=4.0\L^2$ and the lower one is for $m^2_g=4.84\L^2$.}
\label{f1}
\end{figure}
\begin{figure}[htb]
\epsfxsize=0.8\textwidth
\begin{center}
\leavevmode
\epsfbox{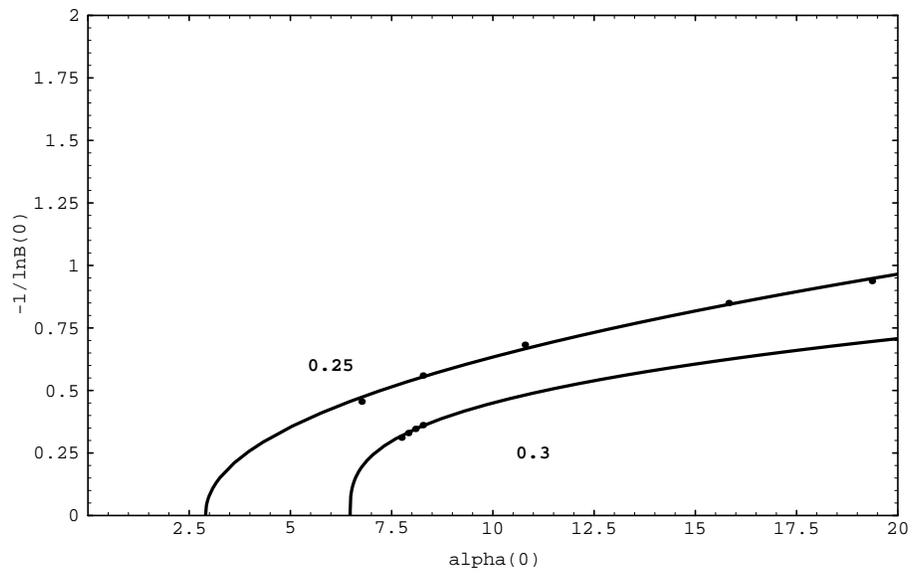}
\end{center}
\caption{The same as Fig.1 for the propagator
of Eq.(\ref{propst}), with $\m^2=0.25\L^2$ (upper curve)
and $\m^2=0.30\L^2$ (lower curve).}
\label{f2}
\end{figure}
\begin{figure}[htb]
\epsfxsize=0.8\textwidth
\begin{center}
\leavevmode
\epsfbox{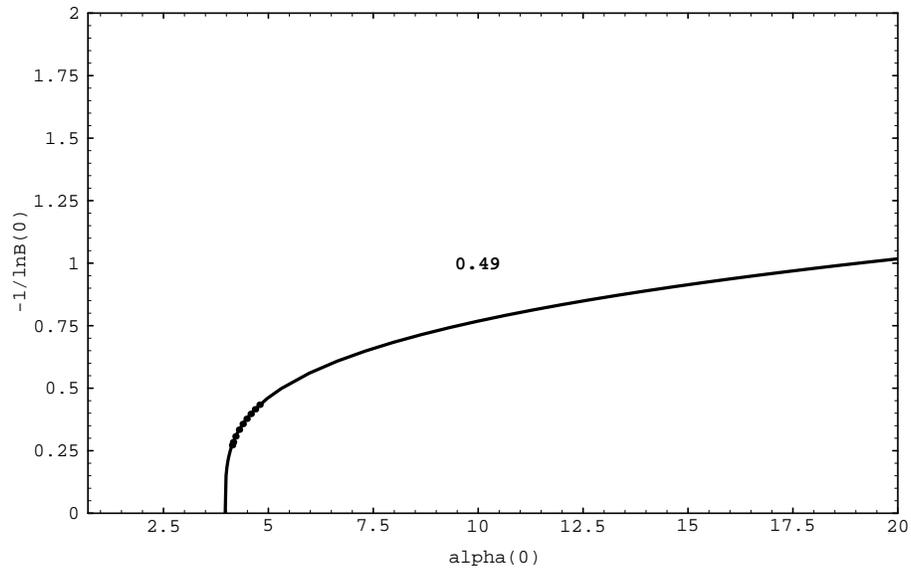}
\end{center}
\caption{The same as Fig.2 for the propagator
of Eq.(\ref{propl}), with $m^2_g=0.49\L^2$.}
\label{f3}
\end{figure}

\end{document}